\documentclass[sigconf]{acmart}

\AtBeginDocument{%
  \providecommand\BibTeX{{%
    \normalfont B\kern-0.5em{\scshape i\kern-0.25em b}\kern-0.8em\TeX}}}

\copyrightyear{2023}
\acmYear{2023}
\setcopyright{acmlicensed}\acmConference[CHI '23]{Proceedings of the 2023
CHI Conference on Human Factors in Computing Systems}{April 23--28,
2023}{Hamburg, Germany}
\acmBooktitle{Proceedings of the 2023 CHI Conference on Human Factors in
Computing Systems (CHI '23), April 23--28, 2023, Hamburg, Germany}
\acmPrice{15.00}
\acmDOI{10.1145/3544548.3581041}
\acmISBN{978-1-4503-9421-5/23/04}

\usepackage{multirow}

\begin{document}

\title[MR.Brick]{MR.Brick: Designing A Remote Mixed-reality Educational Game System for Promoting Children's Social \& Collaborative Skills}


\author{Yudan Wu}
\affiliation{%
  \institution{Institute for AI Industry Research, Tsinghua University}
  \city{Beijing}
  \country{China}}

\author{Shanhe You}
\affiliation{%
  \institution{Institute for AI Industry Research, Tsinghua University}
  \city{Beijing}
  \country{China}}

\author{Zixuan Guo}
\affiliation{%
  \institution{Institute for AI Industry Research, Tsinghua University}
  \city{Beijing}
  \country{China}}

\author{Xiangyang Li}
\affiliation{%
  \institution{Institute for AI Industry Research, Tsinghua University}
  \city{Beijing}
  \country{China}}

\author{Guyue Zhou}
\affiliation{%
  \institution{Institute for AI Industry Research, Tsinghua University}
  \city{Beijing}
  \country{China}}

\author{Jiangtao Gong}
\authornote{Corresponding Author}
\affiliation{%
  \institution{Institute for AI Industry Research, Tsinghua University}
  \city{Beijing}
  \country{China}}
\email{gongjiangtao2@gmail.com}

\renewcommand{\shortauthors}{Wu et al.}

\begin{abstract}
Children are one of the groups most influenced by COVID-19-related social distancing, and a lack of contact with peers can limit their opportunities to develop social and collaborative skills. However, remote socialization and collaboration as an alternative approach is still a great challenge for children. This paper presents MR.Brick, a Mixed Reality (MR) educational game system that helps children adapt to remote collaboration. A controlled experimental study involving 24 children aged six to ten was conducted to compare MR.Brick with the traditional video game by measuring their social and collaborative skills and analyzing their multi-modal playing behaviours. The results showed that MR.Brick was more conducive to children's remote collaboration experience than the traditional video game. Given the lack of training systems designed for children to collaborate remotely, this study may inspire interaction design and educational research in related fields.

\end{abstract}



\begin{teaserfigure}
\centering
  \includegraphics[width=\textwidth]{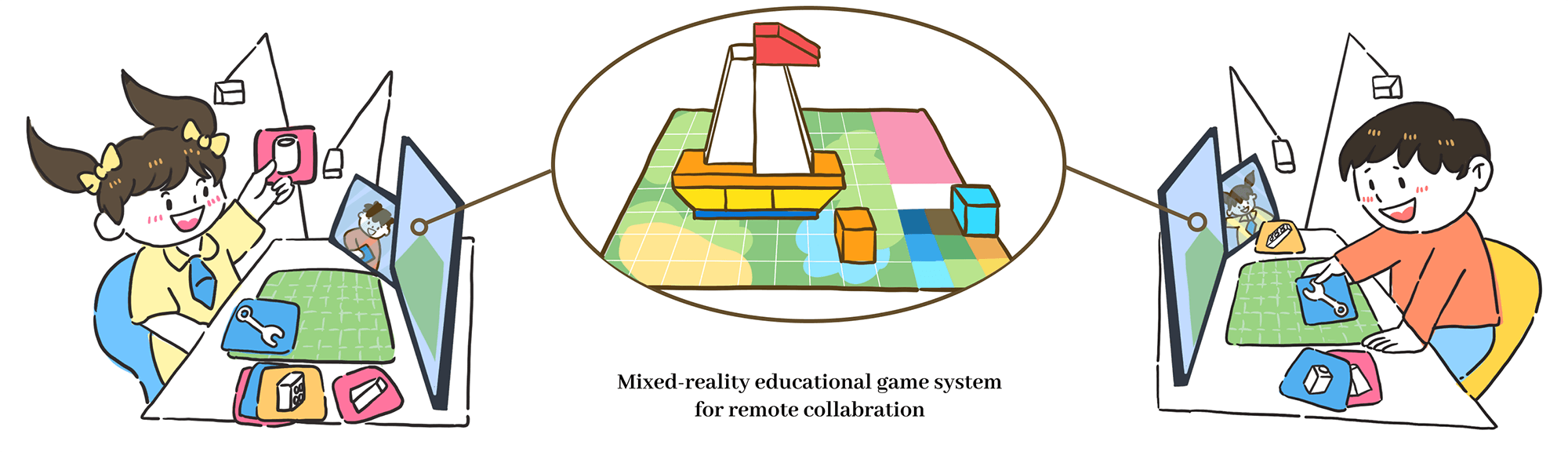}
  \caption{Concept design of MR.Brick: children can play the collaborative game with peers in their own homes.}
\end{teaserfigure}
\maketitle

\section{Introduction}
 The characteristics of human society determine that social interaction and collaboration are basic needs for human survival~\cite{editorial2018cooperative}. Individuals who lack social connection are more likely to suffer from mental health issues such as loneliness, anxiety, and even depression \cite{loades2020rapid}. Compared with adults, children and adolescents are more vulnerable to the negative effects of reducing social interaction due to immature cognitive and emotional regulation systems \cite{zhou2020managing}.

The COVID-19 pandemic and the social distancing measures implemented in many countries have forced the population into isolation and eliminated face-to-face social interaction. 
Following COVID-19, online classes and meetings have become a trend for remote socialization and collaboration. Although children are accustomed to using various devices to access the Internet, remote collaboration is still a big challenge for them. Evidence suggested that children born during the pandemic have significantly lower language, motor and overall cognitive abilities than those born before the pandemic~\cite{deoni2021impact}. Therefore, how to help children reduce or overcome the detrimental impacts of social isolation should be a major concern.

However, there is currently a lack of systems that support remote collaboration for children. Most systems are aimed at adult-child collaboration, and rarely involve child-child collaboration \cite{silva2014opportunities, yarosh2013almost}. Furthermore, some research on collaborative applications for children has limitations, such as being too challenging for young children or failing to consider the impact of remote collaboration \cite{galdo2022pair, orwin2015using, angelia2015design}. In our current age, the lack of techniques to help children learn and adapt to remote collaboration is detrimental to children's development.

In this study, we propose MR.Brick, a remote educational game system based on mixed reality (MR) and tangible interaction technology, helps children increase their online socialization and develop social and collaborative skills. We designed MR.Brick based on multiple cognition and learning theory to improve system usability and reduce barriers for children to collaborate remotely. By conducting a controlled experimental study with 24 children aged six to ten, we found MR.Brick can significantly improve their social and collaborative skills. Furthermore, when compared to traditional video games, children showed more positive emotional and behavioural engagement with the MR game.

Thus, the contributions of this paper are concluded as follows: i) MR.Brick, a novel immersive remote collaborative educational game system based on MR and tangible technology; ii) design guidelines for the remote collaborative system for children based on cognition and learning theory; iii) a controlled experimental study to evaluate MR.Brick by measuring children's pre-/post-social and  collaborative skills as well as system usability scales; iv) comparing MR.Brick with traditional video game through multi-model engagement and collaborative behaviour analysis; v) insights into remote collaborative system design for children.

\section{Related Work}
 
\subsection{Social and collaborative skills}

Collaborative skills, as an important component of social skills, refer to learning how to successfully collaborate with others based on equality to reach common goals~\cite{topping2017effective}. Collaboration is a key facilitator of cognitive development in early childhood \cite{sills2016role}. Peer interaction in collaboration is an external process by which individuals are exposed to opinions different to their own, and verbal communication results in cognitive restructuring \cite{piaget1977development}. Previous studies have confirmed that collaboration benefits children's problem-solving \cite{fawcett2005effect} and rule-based thinking \cite{gabbert1986cooperative}. 

Developing social and collaborative skills in childhood is of great significance to individuals regarding mental health, interpersonal communication, and academic performance  \cite{li2013cooperative}. Children who have been trained and master social and collaborative skills often show more positive attitudes, and are better able to establish relationships with others \cite{piaget2008psychology}. In terms of academic performance, children with these skills are more likely to benefit from cooperative learning and excel academically.

Social life is necessary for children to develop social and collaborative skills, as developing these skills is a learned behaviour that involves initiating, responding, and interacting with others \cite{selimovic2018development}. By limiting children's social activities, the lockdown prevents them from developing social and collaborative skills. Therefore, there is an urgent need to investigate how to help children collaborate remotely through a digital connection. In the proposed game, we try to use mixed reality design to enable children to build blocks together and encourage them to cooperate, communicate and practice.

\subsection{Remote collaboration for children}

During lockdowns induced by pandemic illnesses, remote socialising such as learning, communication, and collaboration has become increasingly crucial for students~\cite{odenwald2020tabletop,yuan2021tabletop,galdo2022pair,gui2021teacher}, with some studies specifically discussing the opportunities and challenges of remote collaboration ~\cite{galdo2022pair,gui2021teacher,yuan2021tabletop}. For example, Galdo et al.~\cite{galdo2022pair} reported positive feedback on students' perceptions of remote collaboration, with some mentioning they felt more successful because they collaborated with their partner, suggesting the possibilities of remote collaboration as a new style of working for young learners. However, research on distant synchronous cooperation among youngsters is still in its infancy. First, some current remote collaboration platforms for children emphasize asynchronous collaboration on long-term projects, such as RALfie~\cite{orwin2015using} and Scratch~\cite{aragon2009tale}. They are more aimed at promoting creative content than helping children socialize and collaborate remotely \cite{orwin2015using,aragon2009tale}. Second, most remote collaboration systems, such as Tabletop Teleporter\cite{odenwald2020tabletop} and Sharetable~\cite{yarosh2013almost}, mainly serve child-adult (such as parents or teachers) collaboration, and only a few focus on child-child collaboration. Angelia et al.~\cite{angelia2015design} proposed a game involving remote collaboration among children, but lacked a systematic description and specific assessment of remote collaboration.

Overall, there is a dearth of remote synchronous collaboration systems intended for child-child collaboration. Children as users differ substantially from other user groups due to their inherent traits. Due to their cognitive and emotional limitations, long-distance communication is challenging for the majority of youngsters~\cite{yarosh2013almost}. Moreover, children are in the stage of developing socialization \cite{stafford2003communication}, which means they have limited attentional resources and incentive to distantly interact~\cite{ballagas2009family}. To fill this research gap, we present MR.Brick, a mixed reality brick-building game to support remote child-child collaboration. 

\subsection{Mixed Reality and Tangible Technology}
Tangible user interface (TUI) blends digital information with physical embodiment in a physical setting. Its design goal is to extend the physical affordances of the products seamlessly into the digital world~\cite{fitzmaurice1995bricks,ishii2007tangible,ullmer2000emerging,gong2014paperlego,gao2022learning}.
Numerous previous studies have demonstrated the possibility and good effects of combining Mixed Reality (MR) and tangible user interface (TUI) to improve children's abilities (e.g. social interaction~\cite{lee2018create}, collaboration~\cite{campos2011fostering}, problem-solving~\cite{lorusso2018giok}, storytelling~\cite{glenn2020storymakar,bai2015exploring}, math learning~\cite{kang2020armath,pontual2018tangible}, and prototyping~\cite{kelly2018arcadia,glenn2020storymakar}). 

MR, including Virtual reality (VR) technology and Augmented reality (AR) technology, has made significant advances over recent years. 
Researchers have found that immersive VR has an advantage over desktop systems when the tasks involved “complex, 3D, and dynamic” content~\cite{mikropoulos2011educational}. 
Some work explored augmenting teachers' views to assist them with classroom routines (e.g. evaluating student's performance~\cite{holstein2018classroom}).
Overall, AR can support pedagogical processes (e.g., providing scaffolding to students) and promote students' engagement~\cite{aslan2019investigating,gong2021holoboard,lu2021design,lu2022chordar,gong2021holoboard2}.

Besides, TUI can facilitate children's play, learning, exploration, and reflection by integrating their visual, auditory and tactile sensory impressions \cite{pontual2018tangible}.The introduction of tangible technology to children's learning has the following advantages: (1) Easy to learn: TUI is a natural interface that requires little cognitive effort to learn; (2) Provide a unique interactive experience: TUI offers the user an alternative way to interact with the computing environment; (3)Attract users continuously: TUI can continuously present the user's interests object, thus supporting Trial-And-Error activity~\cite{xu2005tangible}. Based on the above, we assumed that tangible technology has the potential to help children learn knowledge and improve their overall quality. 

Thus, we implemented children's remote collaboration through the combination of MR technology and TUI in this study.

\subsection{Design considerations based on multiple theories}
In order to make our system better enable remote cooperation and fulfil the needs of children, in this section, we reviewed related theories, including cognitive science, developmental psychology, and cognitive load theory. Moreover, we concluded six design guidelines for further system design.

\subsubsection{\textbf{Shared attention} is an essential precursor to collaboration\cite{antle2013getting}} Users cannot consult meaningfully to reach a common conclusion and understanding unless they notice what each other is doing.
Yuan et al. demonstrated that adopting hybrid setup (e.g., webcam, tripod, mirror board) in a shared-task space can better assist in engaging players as well as supporting their remote collaboration and socialization during the tabletop game session~\cite{yuan2021tabletop}. Similarly, based on theories of CSCL~\cite{dillenbourg1999collaborative}, Antle et al. suggested that tangible user interface designers should create configurations where players are able to observe others' movements and attention~\cite{antle2013getting}. Based on above, we developed our design guideline 1: 

\begin{quote}
Design Guideline 1: Create a remote mixed reality shared-task space for multiple players, and visualize their real-time modification.
\end{quote}

\subsubsection{\textbf{Non-verbal cues} (such as eye gaze, facial expression, vocal tone, and body language) play an important role in facilitating communication~\cite{mast2007importance,tepper1978verbal,archer1977words,sundaram2000role}, even greater than verbal ones~\cite{tepper1978verbal}.}
However, these cues are often absent in remote games. 
The insufficient awareness of the environment and other players could bring negative effects on players' collaboration experiences~\cite{yuan2021tabletop}. Yuan et al. suggested designers to set a shared communication space(e.g., videochat, audiochat, text messages) to support collaboration, discuss strategies, as well as participate in social conversations. Therefore, here comes our design guideline 2: 

\begin{quote}
Design Guideline 2: Setup additional communication space for gaming and socializing.
\end{quote}

\subsubsection{\textbf{Image schemas} are mental structures based on patterns of experience of high frequency of occurrence~\cite{johnson2013body}}.
That is, the simple and basic cognitive structure that people acquire repeatedly in the process of interacting with the objective world in daily life.
Antle et al. proposed that these schemas can be used to design input actions that users often perform unconsciously, or are easy to learn~\cite{antle2013getting}. There were also evidences that leveraging image schemas had usability advantages~\cite{antle2008playing,bakker2010moso,bakker2012embodied}, allowing both child and adult to focus more on using than learning to use a system, which is consistent with our design goal. Thus, we formulated our design guideline 3: 

\begin{quote}
Design Guideline 3: Using image schemas to design input space.
\end{quote}

\subsubsection{\textbf{"Jigsaw" scripts} are often used as a way to create positive interdependence in a collaborative situation in CSCL~\cite{antle2013getting}}
In TUI systems, a "jigsaw" script can be enacted through both physical objects and learning activity~\cite{antle2013getting}.
For example, imposing restrictions on information each player needed to
accomplish a collaborative task~\cite{miyake2001complex}, distributing roles and controls\cite{antle2013getting}, or by creating a constrained input system that requires each player to take a specific action in order~\cite{hornecker2005design,antle2013getting}. 
Interestingly, we found a similar concept demonstrated in the field of psychotherapy. 

\emph{\textbf{LEGO therapy} is developed as a therapeutic method to improve the social skills of children with ASD~\cite{LEGOff2004use}.}
The mechanism behind LEGO therapy is to create interaction opportunities and motivate children to work together by building in pairs or small groups with a social division of labour~\cite{owens2008LEGO}. It requires players to follow a particular order and take specific action to complete the LEGO build and obtains improvements on children's social competence~\cite{LEGOff2004use,owens2008LEGO,LEGOff2006long}. Based on the above, we developed our design guideline 4 : 
\begin{quote}
Design Guideline 4: Create a constrained input system and distribute roles, information and controls.
\end{quote}

\subsubsection{\textbf{Scaffolding} refers to the support in the purpose of providing learners with the necessary assistance to enable them to complete tasks and develop understandings that they cannot handle on their own~\cite{hammond2005scaffolding}}
Its theoretical basis lies within the framework of Vygotsky's~\cite{hammond2005scaffolding}, 
who emphasized that appropriate social support is essential for children to learn~\cite{vygotsky1978mind,hourcade2008interaction}. Analogously, Giusti et al. stressed the importance of the facilitator's engagement and power in educational tabletop games~\cite{giusti2011dimensions}. The facilitator is essential for moulding children’s experience, because they can control the pace of the activity, influence the dynamics between the children, and help them to achieve an expected performance~\cite{giusti2011dimensions}. Various prior works had suggestions and applications of building a guidance agent to provide instructions and emotional support~\cite{gossen2012search,becsevli2019mar,africano2004designing,hopkins2011avatar}. Based on above, we enacted our design guideline 5: 
\begin{quote}
Design Guideline 5: Build an agent to achieve high support.
\end{quote}
\subsubsection{\textbf{Maturation} is one of the four major factors that Piaget thought affected development~\cite{piaget2008psychology}.} Children’s physical maturation limits what and how they are able to learn~\cite{hourcade2008interaction}. 
According to children’s developmental psychology~\cite{piaget2008psychology}, our participants lie within the concrete operations stage (7–11-year old), which is more likely to appreciate someone else’s perspective, letting them to better work in teams and as design partners with adults~\cite{hourcade2008interaction}. However, their limited cognitive and motor abilities bring challenges to the design of AR applications~\cite{radu2012using}. We will address these limitations based on categories from Radu et al.~\cite{radu2012using} below:

\textbf{-Gross Motor Skills and Endurance.}
Games are reported as straining after 10 minutes even  when users play on a table surface~\cite{correa2007genvirtual}. 

\textbf{-Hand Eye Coordination.}
Children find it difficult when the direction of hand movement does not match the direction of eye movement~\cite{dunser2007lessons}. 

\textbf{-Fine Motor Skills.}
This skill becomes strained when children have to move precisely in small areas and when actions need to be done under time limits~\cite{correa2007genvirtual}. 

\textbf{-Spatial visualization and spatial perception.} Young children can recognize objects and their relative sizes, but have trouble estimating distances. Moreover, children before 8 years old also have trouble with mental rotations~\cite{rosser1994cognitive}. 

\textbf{-Divided attention.} Until  about  8  years  old,  children  can  only  focus on one item/activity at a time~\cite{rosser1994cognitive}. The following situations of AR may require a child's ability of divided attention~\cite{radu2012using}: 1 )Input and output are in different Spaces; 2) Virtual content occludes physical action; 3) Observe AR game and attend to instructions at the same time; 4) Advanced game design may require children to focus on multiple items at once. 

\textbf{-Selective and Executive Attention.} 
Studies have shown that children have trouble controlling attention. For example, they may have trouble concentrating and become easily distracted~\cite{rosser1994cognitive}. 

Being aware of what most children at this age are able to accomplish can provide designers with useful guidelines~\cite{hourcade2008interaction}.
Therefore, we developed our design guideline 6:
\begin{quote}
    Design guideline 6: Design a children-friendly manipulation to meet the age group's needs.
\end{quote}

\section{Design}
MR.Brick is designed as a collaborative mixed reality building game in which two child players work remotely. 
The players construct, move, and transfer virtual bricks in 3D virtual space by manipulating the tangible tool.
They can communicate verbally and nonverbally at all times while playing, just as if they were face-to-face with LEGO building.

\subsection{Designing MR.Brick based on six design guidelines}
As we concluded in section 2.4, we reviewed related theories and produced six design guidelines to make our system better enable remote cooperation and fulfil the needs of children. In this session, we described how they are utilized inside our system.

\subsubsection{\textbf{Design Guideline 1: Create a remote mixed reality shared-task space for multiple players, and visualize their real-time modification.}}
We utilized a tangible board as a real-time collaboration space for virtual shared activities. The physical board, which is covered in AR recognition patterns, is transferred onto the virtual playroom on the display. In addition, we included 16 tangible cards with AR recognition patterns as manipulation tools for players, allowing them to select, create, and move various virtual blocks. During the game, as one player manipulated on one side, the other player on the opposite side could see the change instantaneously reflected on the screen, including hand positions in real-time. This environment is similar to Minecraft but with augmented content, or a mixed-reality sandbox game. Thus, our technology enables participants to watch one other's on-screen motions in real-time.
\begin{figure*}[h]
    \centering
    \includegraphics[width=\textwidth]{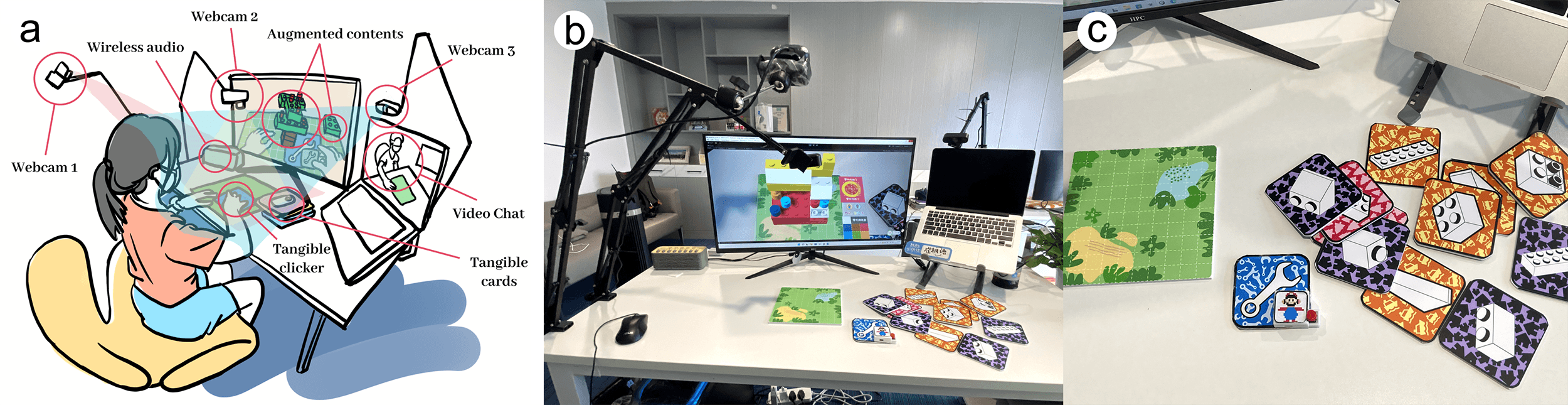}
    \caption{A remote mixed reality shared-task space for children. a) the sketch diagram of the physical environment; b) the final implementation of the physical environment; c) the tangible toolkit.}
    \label{fig:environment}
\end{figure*}

\subsubsection{\textbf{Design Guideline 2: Setup additional communication space for gaming and socializing.}}

We leveraged an additional video chat platform for each participant, consisting of an additional screen and an external webcam. To ensure that all actions could be captured, the webcam was positioned to capture the player's facial expressions and hand movements, as well as the tangible toolkit on the table. We preserved the players' knowledge and information as much as possible through this method.

\subsubsection{\textbf{Design Guideline 3: Using image schemas to design input space.}}

In our design, if players want to better observe their collaborative configuration from different perspectives, they just need to simply rotate the tangible board. As for the manipulative tools, players must move the tangible cards spatially(that means, to move up-down, back-forward and rotate) in the virtual playground to achieve the expected response, analogous to the motions required to complete practical brick-building tasks.
This type of unconscious input action allows children to focus more on the game and less on their manual operation.
Moreover, to accommodate unavoidable interactions with digital content throughout the game, we incorporated a wireless clicker into a 3D-printed box of the same size as the physical cards and adorned with AR recognition patterns.
Compared with interacting with conventional pointing input devices (e.g., mouse, trackballs, joysticks, keyboards), this technology enabled seamless implicit interactions without the need to switch between tangible tools and extra devices.

\subsubsection{\textbf{Design Guideline 4: Create a constrained input system and distribute roles, information and controls.}}
\begin{figure*}
    \centering
    \includegraphics[width=\textwidth]{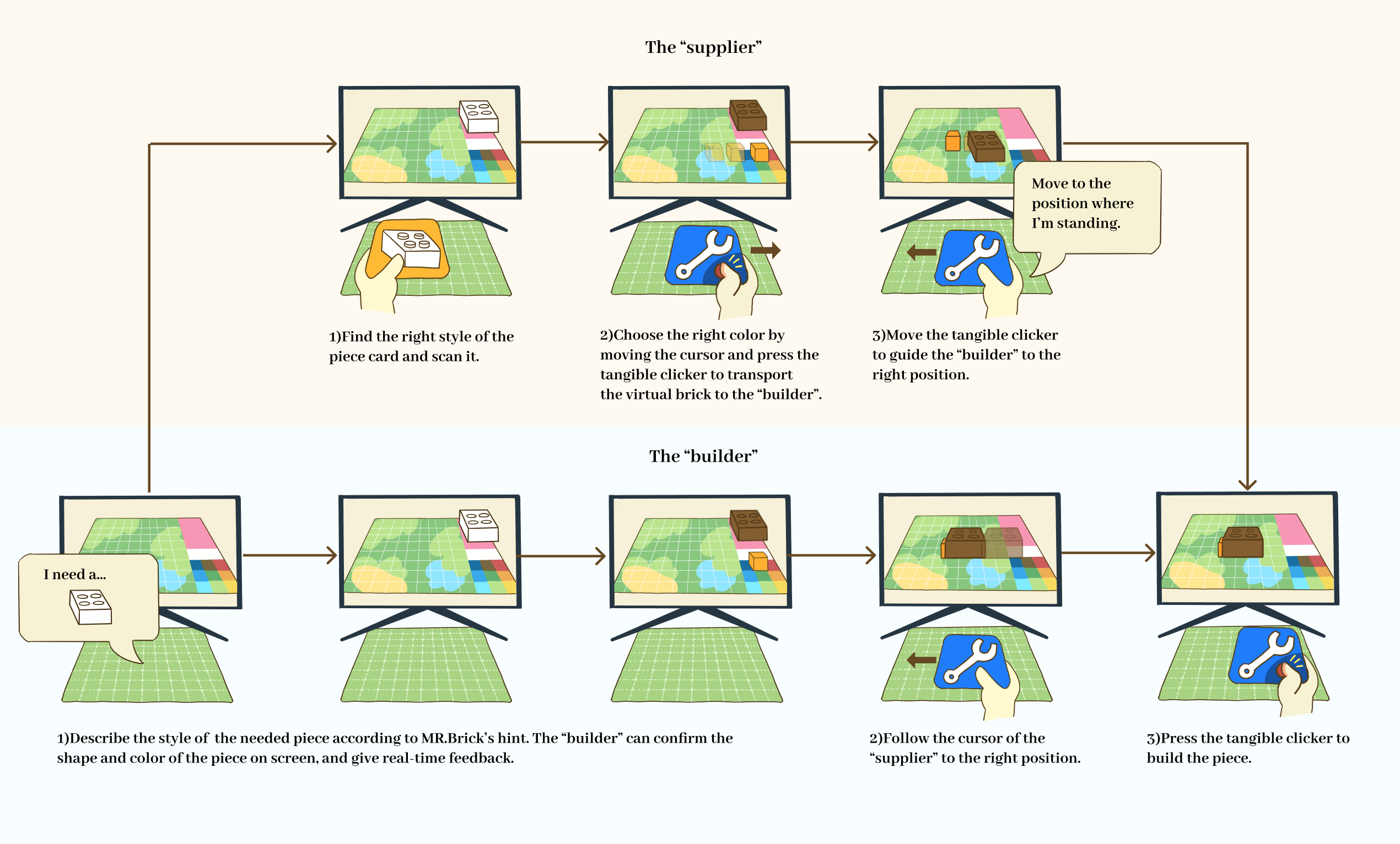}
    \caption{Game rules and interaction design of MR.Brick: the ''supplier'' and ''builder'' need to communicate and collaborate closely with each other.}
    \label{fig:rule}
\end{figure*}
As shown in Fig.\ref{fig:rule}, our design adheres to the fundamental rule of LEGO therapy. One child is the "supplier" (collects the necessary pieces), while the other is the "builder" (builds the pieces together). The traditional LEGO therapy project includes a third role, the "engineer," but after careful study, we deleted this function and united the character with the "builder." Players would assume their assigned role for a single gaming session before switching roles. To transfer the pieces, the "supplier" must: 1) choose the correct shape of the piece; 2) apply the correct colour to the piece; 3) press the tangible clicker. The piece will then be visible on the "builder" side of the screen. After obtaining the correct piece, the "builder" must: 1) place the brick in the hint position; 2) click the tangible clicker to construct. The "supplier" could only transport the next required component once the "builder" completes construction. This well-organized design of game rules for turn-taking accomplished our goal of establishing a collaborative, interdependent environment.

Moreover, according to Antle et al., in order to create positive interdependence, designers can consider supporting private usage of physical objects and using movable digital representations for public used objects\cite{antle2013getting}. In our design, we adopted a set of tangible tools for multiple usages of different roles. Due to the "supplier's" task of locating and colouring the correct brick pieces, 16 cards with unique recognition patterns for each piece were created. As for the mission of the "builder"'s building and the "supplier"'s transporting, we built a card with an embedded wireless clicker and a hammer-shaped recognition pattern, one for each player. Apart from the private usage of tangible tools, all of our objects intended for communal use were depicted digitally.
    
\subsubsection{\textbf{Design Guideline 5: Build an agent to achieve high support.}}

An agent named Mr.Brick was created to facilitate and foster collaboration and communication among children, capture their failures, and provide accurate guidance. For instance, when the "builder" has a new mission, Mr.Brick will appear in a pop-up window and provide an image of the required item, along with a voice notice guiding and encouraging the "builder" to direct the "supplier". When the "supplier" collects the incorrect piece, Mr.Brick will appear with a bubble text box telling them of their error and encouraging them to try again.

\subsubsection{\textbf{Design Guideline 6: Design a children-friendly manipulation to meet the age group's needs.}}
We will explain the corresponding design based on the categories mentioned in our related work session. 

\textbf{-Gross Motor Skills and Endurance.}
Each round of the game should run no more than ten minutes and no less than five minutes.. 

\textbf{-Hand Eye Coordination.}
The webcam we utilized as input to show on the screen was positioned from the top of the child's head to the tangible board, so that the digital material shared the same view as the children's observation. We avoided the mirroring issue this way.

\textbf{-Fine Motor Skills.}
The virtual bricks are instantly absorbed into their respective locations, eliminating the need for precise card motions.

\textbf{-Spatial visualization and spatial perception.} A virtual cube describing the real object tracking position and a second cube matching to the personal moving cursor representation in the game field has been implemented. These two squares are nearly identical, yet their angles differ. This is intended to help children comprehend how tool movement in the hand corresponds to the movement trajectory of the virtual item. Together with the tangible board that can be rotated 360 degrees to see the material from multiple perspectives, children may learn to comprehend and utilize 3D virtual content with greater ease.

\textbf{-Divided attention.} We matched the physical and virtual content types so that children may rapidly comprehend the relationship between the physical input and virtual output space. Additionally, we made the input hardware equipment very easy to operate — even without looking — and reduced the frequency of sight transfers between the two places by a large amount. We also altered the virtual field's opacity so that the player's hand movements are clearly seen on the screen and are not obscured by the virtual content. In order to prevent children from ignoring the instructions owing to the allure of the game, we made each step appear as a popover in the most visible area of the game screen (which cannot be interacted with) and lasted for three seconds so that children could pay attention. To prevent children from being distracted from AR tracking, we modified a number of design aspects, including: 1) After successful recognition, the "supplier's" delivered brick is fixed on the portal area by default, and there is no need to keep the recognition card within the camera's range; 2) During the same period of the game, only the tangible board and one tangible card need to be recognized.

\textbf{-Selective and Executive Attention.} 
In order to relieve the burden of attention control, we devised the rules of the game so that sending and constructing occur sequentially, i.e., the "builder" cannot begin assembling bricks until the "supply" has remotely sent the appropriate piece. This forced the children to stop focusing on their own game and instead observe what other participants were doing.

\subsection{Technical Details}
Following our design guideline, refined through a series of pilot studies, we developed MR. Brick, a remote mixed-reality educational game. The development of the game was separated into "virtual" and "real" phases. The "virtual" component was the Unity3D-powered video game, while the "real" component was the tangible toolkit, including the tangible board (game map), tangible cards (bricks), and tangible clickers (input devices).

\begin{figure}[h]
    \centering
    \includegraphics[width=0.5\textwidth]{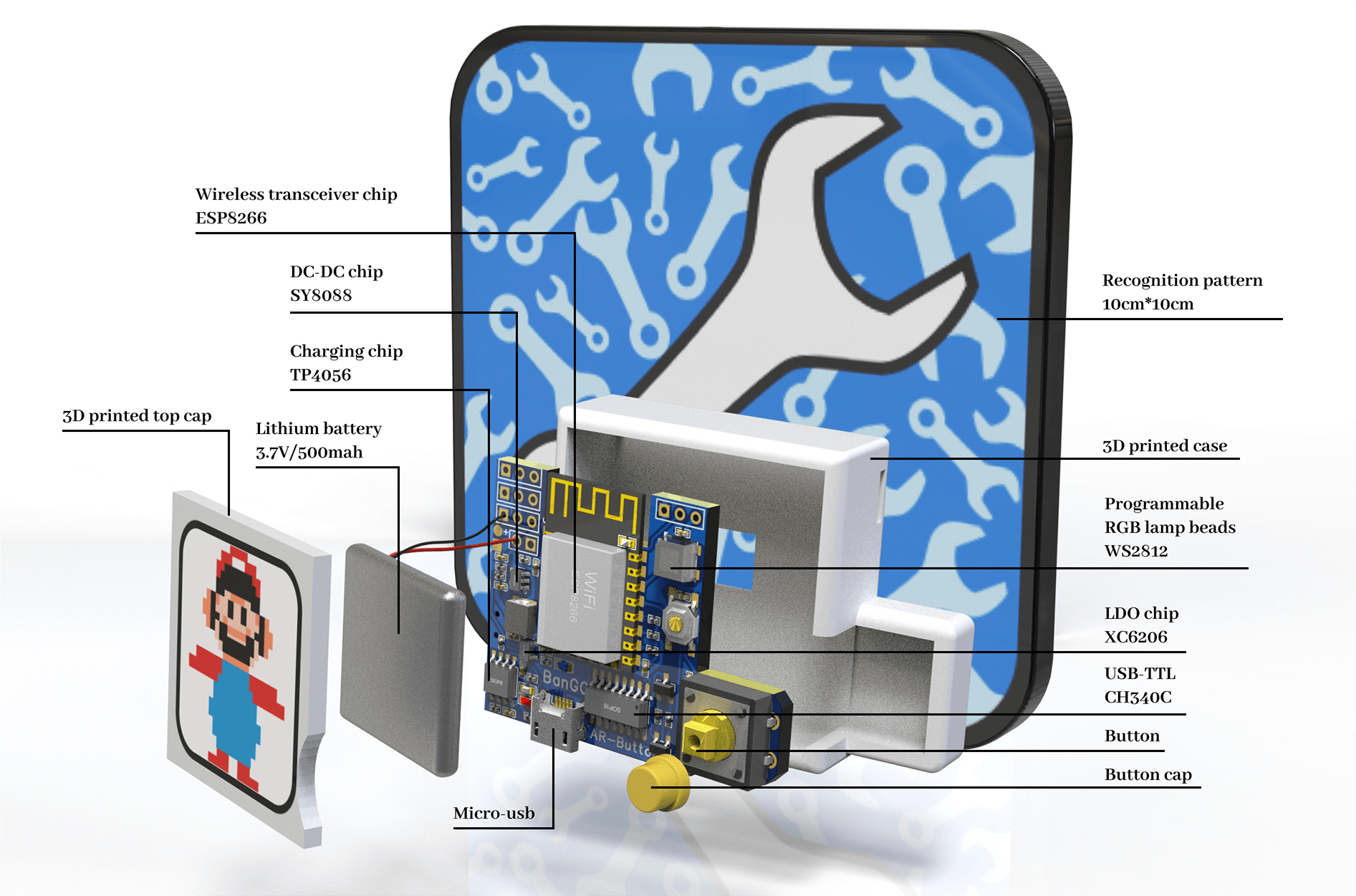}
    \caption{Hardware components of tangible toolkit in MR.Brick system.}
    \label{fig:hardware}
\end{figure}

\subsubsection{\textbf{Hardware}}
In the game, each kid has their own set of tangible toolkits, which are explained in further detail below.
\begin{enumerate}
    \item a 20cm*20cm PVC tangible board with recognition pattern; 
    \item 16 10cm*10cm PVC tangible cards with recognition pattern; 
    \item a tangible clicker (Figure \ref{fig:hardware}) with 3D printed case and embedded circuit board (consists of a wireless transceiver chip ESP8266 for WiFi connection, a DC-DC chip SY8088, a USB-TTL CH340C, a LDO chip XC6206 and a programmable RGB lamp beads WS2812);
\end{enumerate}

Our hardware system also includes of the following devices to support mixed-reality installation and video online conferences:
\begin{enumerate}
    \item 3 HIKVISION E14a 2K cameras; 
    \item 3 Aluminium alloy table top bracket; 
    \item a 32-inch 75Hz IPS hard screen display; 
    \item 2 laptop for video online meetings.
\end{enumerate}

We installed two high-resolution cameras for toolkit detection, with camera 1 recording from a greater distance at an oblique angle, and camera 2 situated closer to the actual toolkits from above at a vertical angle.
Children can observe the view from camera 1, which simulates their first-person perspective, facilitating their comprehension of the relationship between augmented information and reality in studies. However, using a single camera at an oblique angle is susceptible to linear perspective issues, which could lead to inaccurate distance and position calculations. For more precise location computations, we added camera 2 at a vertical angle to the table. By utilizing the position data captured by camera 2, the displayed image may be guaranteed to be correct and easy to understand.

\subsubsection{\textbf{Software}}
We developed this game using the Unity3D engine because of its low cost, easy  updating, and strong expandability. Three well-developed and supported packages were used in our game, including the network package Fusion, the Augmented Reality package Vuforia, and the unity client duplication tool Parrel Sync.

\textbf{Fusion} network package was used to provide a robust and low-latency game network in the game. In the implemented Host and Client mode, the first participant to start the game will become the "Host," which is both the game's server and client. Later-joining players will automatically search for and join the game as clients. This mode only allows the server to modify the networked, shared game object, ensuring game consistency and reducing the requirement for network communication. This mode is also child-friendly, as the game will be formed automatically upon program launch.

\textbf{Vuforia} package is the augmented reality package for image identification and tracking.  In our technique, we use the tangible board as the centre of the world, the  relative locations of other objects  can then be computed and scaled for proper display. To achieve the correct positioning that children can understand, we carefully alter the size and proportion of the tangible board and tangible cards for more precise and realistic performance.

\subsection{Pilot Study and Design Iteration}
\subsubsection{MR game version}
After three rounds of pilot research involving six people between the ages of six and nine, we discovered the following issues:
1) each participant had all the necessary information to construct a brick, so communication was not required to complete the game; 2) "engineer" role players may find themselves idle due to a lack of opportunities to connect with virtual content. 3) the challenge of tracking the tactile cards for children in the MR context, and the problem of turning the camera in the standard video game setting; 4) players have trouble verbally guiding one another to make the correct move.

Here are the system modifications we made in response to user feedback and experimenters' observations:

1) limiting the number of players to two; 2) developing a new rule for the game that will be distributed. In the new rules, the 'builder' will not know the position of the brick to build, and the 'supplier' will not know the type and colour of the brick required. This forces the two players to share information and guide each other in order to complete the game; 3) increased the size of the tangible card from 4*4cm to 10*10cm and added more recognition spots on the pattern to improve the recognition accuracy; 4) adjusted the game field's transparency to 50$\%$ to make players' hands and tools visible when manipulating; 5) fixed a few details of the game operation, such as: only the correct bricks can be transferred, and incorrect bricks will only be given a prompt (to avoid an extra functional development of backward retraction, which can be difficult for children at this age to understand, and prevent the chaos caused by children's errors); 6) adjusted the webcam angle to better capture the players' hands and tools; 7) visualized the real-time position of the player's cursor by displaying colored blocks, so that the "supplier" could use their cursor to direct the "builder" to the precise location to construct; 8) produced instructional videos and revised the order and content of the experimenter's guide.

\subsubsection{Traditional video game version}
Together with the MR version, the standard video game version was developed and optimized repeatedly. Two volunteers between the ages of 6 and 9 evaluated its applicability and accessibility in two rounds of pilot testing.
The traditional video game version shared the same game rules, as well as the same graphical design and manual instructions. The main difference is how they interact with the game; one uses a mouse and the other uses tangible cards.
After the preceding rounds, we settled on the version of the system we would use for the experiment.

\section{Experimental study}
In our study, two children collaborated as a group and played different roles to perform game tasks. Traditional video games, such as computer games, represent a currently popular and common remote collaboration experience. To evaluate the quality and effect of remote collaborative games for children, this study set two conditions based on game type:
\begin{itemize}
\item Mixed reality game - Mixed reality game with toolkit and AR technology.
\item Traditional video game - Traditional computer games without toolkits and AR technology.
\end{itemize}
 
We used between-within mixed-method in this study, thus, all children completed game tasks in both conditions. Half of the groups played the MR game first and then the traditional video game, and the other half did the opposite. Children working in pairs were separated into different rooms to play games in both conditions. After the game, children can get prizes including brick toys, snacks, and certificates. Furthermore, this research received IRB approval from our institution.
\begin{figure}[h]
    \centering
    \includegraphics[width=0.5\textwidth]{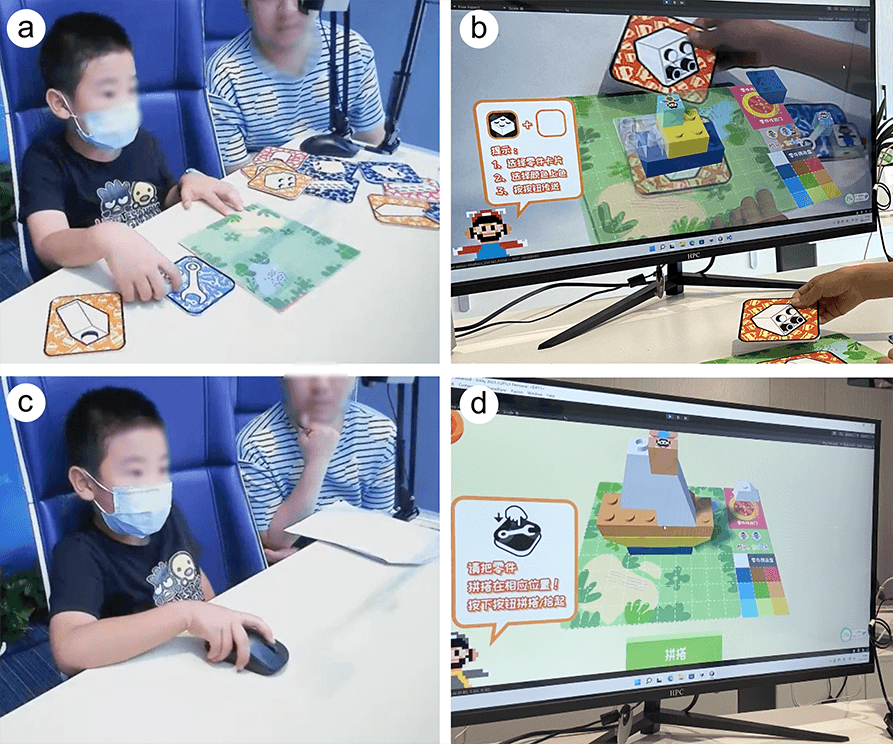}
    \caption{The difference between MR game and traditional video game. a) a boy using tangible tools to play the MR game;  b) the user interface of the MR game; c) a boy using a mouse to play the traditional video game; d) the user interface of the traditional video game.}
    \label{fig:mrvsv}
\end{figure}

\subsection{Participants}
In total, 24 children aged 6 to 10 participated in our study (M = 8.17, SD = 1.07, the demographics information see in Appendix A.1, Table.~\ref{tab:demographics} ). We recruited these participants by distributing an electronic poster to potentially eligible parent groups. The poster includes a general introduction (i.e., a brick-themed remote collaboration game), time, location, and prizes offered. There was also a QR code linking to the online questionnaire on the poster. Parents who are interested in signing up need to fill in the questionnaire (see Appendix A.2). The first section of the questionnaire was the demographic information of the participating children, and we excluded some children who were not in the age range of 6-10. The second section was information related to the experiment, including children's interest and familiarity with brick-building games, children's familiarity with mouse control, children's emotional state during the pandemic, and children's basic social and collaborative skills. Parents' written informed consent was obtained for all children participating in the experiment. 

\subsection{Procedure}
\begin{figure*}[htbp]
    \centering
    \includegraphics[width=\textwidth]{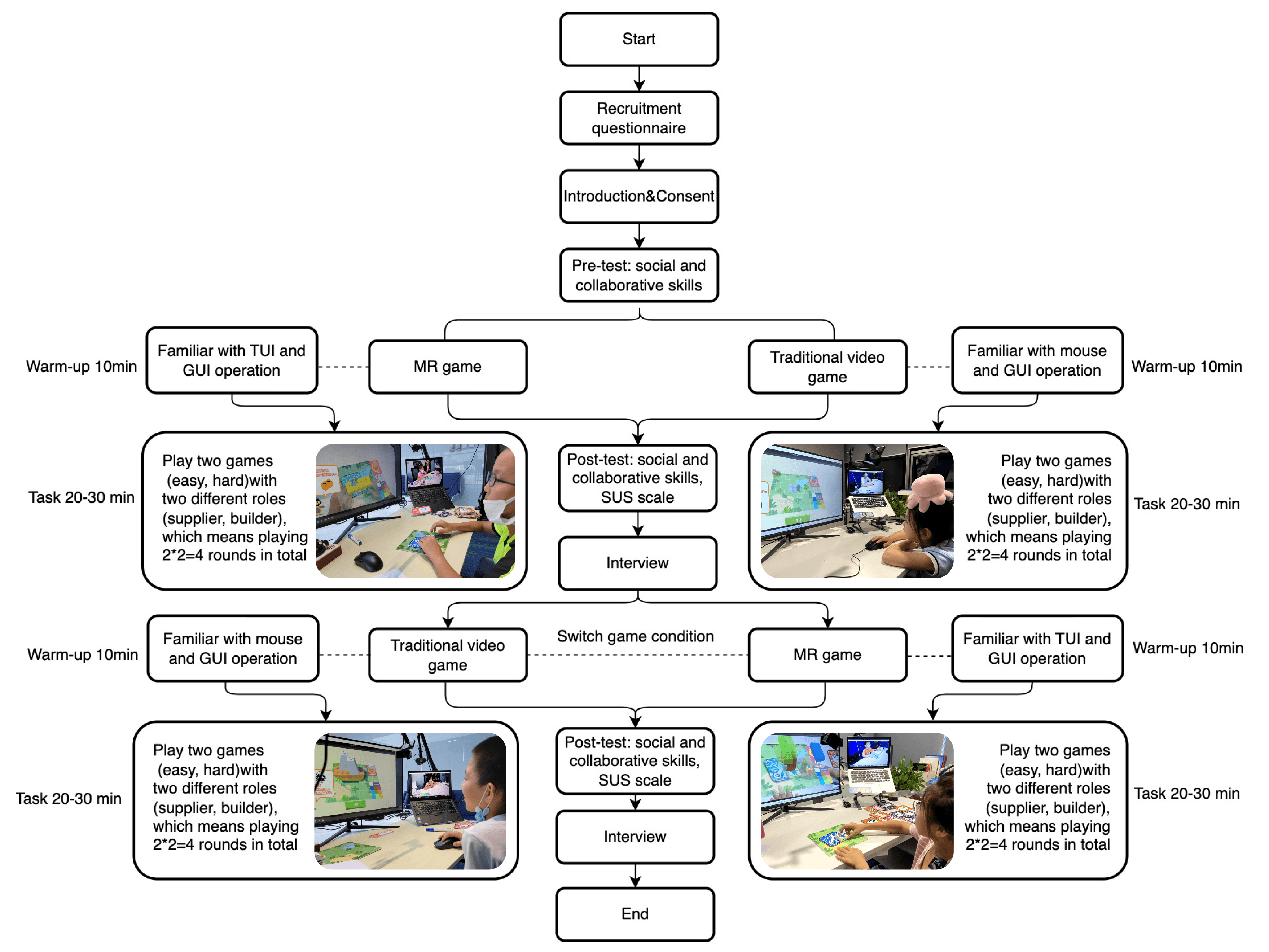}
    \caption{Procedure of between-within mixed-method adopted in this experimental study.}
    \label{fig:procedure}
\end{figure*}

We randomly divided the children into two-person groups based on their availability, ensuring that the age gap between them was as close to 1-2 years as possible. Research shows that cooperation between children of the same age benefits all participants, whereas cooperation between children of different ages only benefits younger children \cite {sills2016role}. After grouping, the experimenters determined the order of their experiments (i.e., the MR game first or the traditional video game first). For each condition, all children were required to complete a warm-up and formal experiment. To ensure the consistency of the experimental process, we prepared an introduction video to explain the rules and related operations. Moreover, each child was paired with an experimenter. The two experimenters had been properly trained to guide the children through a fixed instruction script. During the 10-minute warm-up, children familiarized themselves with the game rules and related operations mainly through the introduction video. The formal experiment lasted 20-30 minutes and required children in a group to complete an easy task followed by a hard task. Each task needs to be completed twice to ensure that the children experienced two roles (i.e., supplier and builder), resulting in two (tasks) x two (roles) = four rounds of the game. The difference between easy and hard tasks was the average number of bricks, 9 for easy and 12 for hard. To keep the children engaged and motivated, the game tasks in the two experimental conditions (i.e., MR game and traditional video game) were different. The experimental process is shown in the Fig.~\ref{fig:procedure}.

\subsection{Measures}
This study used objective and subjective measures, mainly including performance on collaborative tasks, social and collaborative skills, and gaming experience.

\begin{enumerate}
\item \emph{\textbf{Log Data.}} To evaluate the performance on the collaborative task, we recorded the group's time to complete the task and the correct rate of building bricks based on log data.

\item \emph{\textbf{Video Data and Analysis.}} The experiment was completely recorded by cameras, including the video online meeting camera on the laptop and two HIKVISION E14a cameras placed on both sides of the children. Through manual coding, video recordings were primarily utilized to analyze children's engagement and collaborative communication. As indicated in Table \ref{code}, the coding scheme was developed and incorporated from Woodward et al. \cite{woodward2018investigating} and Gong et al. \cite{gong2021holoboard}, and it included emotional engagement, behavioural engagement (nonverbal), and verbal communication. All children were coded and we randomly selected two video clips of them playing different roles, one from the MR game and the other from the traditional video game. Three trained coders independently encoded these videos with an inter-rater Kappa greater than 0.89.

\begin{table*}[htbp]
\caption{Manual coding scheme for engagement and collaborative behaviour}
\centering
\begin{tabular}{lll} 
\hline
Categories                               & Types    & Description                                                                     \\ 
\hline
\multirow{2}{*}{Emotional
  Engagement}  & Positive & Joy: happy, excited, aroused, relaxed, satisfied, etc.                          \\
                                         & Negative & Boredom/Anger: bored, depressed, miserable, annoyed, frustrated, etc.           \\ 
\hline
\multirow{2}{*}{Behavioral
  Engagement} & Positive & Attentive behavior: look at the camera, gesture, show the card,
  etc.          \\
                                         & Negative & Inattentive behavior: lean back, rotate chair,~ look
  away, etc.               \\ 
\hline
\multirow{2}{*}{Verbal Communication}    & Positive & Smooth collaboration: commentary, cheering, patient, encouraging, etc.          \\
                                         & Negative & Frustrated collaboration: critical, complaining, impatient, speechless,
  etc.  \\
\hline
\end{tabular}
\label{code}
\end{table*}

\item \emph{\textbf{Pre-/Post- Scales.}} We prepared different scales for assessing social and collaborative skills and system usability. Before the experiment began, children were tested on social and collaborative skills, and then after each condition of the game, they were given a social and collaboration skills test and a system usability test. We utilized the Collaboration Self-Assessment Tool (CSAT)\cite{ofstedal2009collaboration}, a four-point scale with 11 items, to assess social and collaborative skills. To measure the system usability, we applied an adapted System Usability Scale (SUS) \cite{brooke1996sus} (see in Appendix A.3). We first adjusted the questions to be easier for children to understand, and eliminated two difficult questions: "I thought there was too much inconsistency in this system", and "I found the system very cumbersome to use". Furthermore, we changed the question "I found the various functions in this system were well integrated" to "I think the game is well designed and both roles are fun to play", to better understand their attitudes towards functional integration from the perspective of two roles.

\item \emph{\textbf{Interviews.}} We interviewed all children about their satisfaction and playing experience after they had played in each condition. Our major questions for satisfaction were, "Do you like this game?" and "Which part do you find the most/least fun?" For the gaming experience, we primarily inquired about their collaboration experience, co-presence experience, and possible relief from loneliness. Interviews were fully recorded and transcribed for further data analysis.
\end{enumerate}

\section{Results}
We compared the overall performance of the children in the MR game with the traditional video game in terms of game completion time and accuracy rate. The average time required to complete a game of MR was 382.83 seconds, compared to 382.13 seconds for the traditional video game. The average accuracy rate of children constructing blocks in the MR game was 65.49\%, compared to 61.75\% in the traditional video game. The results suggested that there were no significant differences between the two games in terms of the children's overall performance.

Using the SUS scale, we validated both games' usability. The total SUS score for the MR game was 70.05 (SD = 1.925), whereas the score for the traditional video game was 70.60 (SD = 1.868). Based on the results of the Student's t-test, we found no significant difference in overall usability performance between the two games. However, the two games were borderline statistically significant in terms of willingness to play and ease of use, as seen by the children's stronger inclination to play the MR game more frequently in the future (p=.096, t=1.346).

\subsection{Social and Collaborative Skills}
We conducted Student's t-test to compare the social and collaborative skills pre-test results as the baseline between the MR game and the traditional video game. The results showed that both games could significantly improve children's social and collaborative skills (p<0.01,t=-5.615), but there was no significant difference between the two games (p=.130,t=-1.186). Children's post-test scores (M=50.583, SD=7.38) were significantly higher than their pre-test scores (M=44.250, SD=4.719) after completing the MR game (p<.001,t=-4.253). At the conclusion of the traditional video game, the children's post-test scores(M=51,917, SD=3,655) were significantly higher than their pre-test scores (M=49.250, SD=4.181) (p=0.002, t=-3.499). In addition, among the 11 items of social and collaborative skills scales, we identified 9 items in which children demonstrated significant improvement in both conditions, namely Contribution (p=.055,t=-1.661), Team Support (p=.044,t=-1.781), Problem Solving (p=.008,t=-2.584), Team Dynamics (p<.001,t=-3.892), Motivation/ Participation (p=.002,t=-3.243),Quality of Work (p=.006,t=-2.696),Time Management (p=.001,t=-3.331),Preparedness (p=.077,t=1.476), and Role Flexibility (p<.001,t=-3.921).

\subsection{Engagement and Communication}
According to the Student's t-test results of video coding, children who participated in the MR game had more behavioural engagement, emotional engagement, and verbal communication, as indicated in Table.\ref{Table:InterT}.

\begin{table*}[htbp]
\centering
\caption{Student's t-test results of video coding (* indicates p<0.05, ** indicates p<0.01, *** indicates p<0.001)}
\begin{tabular}{llllll} 
\toprule
Data                                   & Variable & Mix-Reality Game & Traditional Game  &Value    & p      \\
\hline
\multirow{2}{*}{Behavioral Engagement} & Positive & M=14.625 (SD=6.933)**         & M=10.250 (SD=6.476)           & t=2.526  & 0.009  \\
                                       & Negative & M=8.250 (SD=6.739)           & M=14.833 (SD=11.919)**          & t=-3.028 & 0.003 \\
                                       \hline
\multirow{2}{*}{Emotional Engagement}  & Positive & M=7.083 (SD=5.763)**          & M=4.750 (SD=4.571)          & t=5.684  & 0.005  \\
                                       & Negative & M=1.292 (SD=1.367)           & M=2.833 (SD=2.697)*
                                       & t=-3.531 & 0.017   \\
                                       \hline
\multirow{2}{*}{Verbal Communication}     & Positive & M=0.792 (SD=0.884)**        & M=0.417 (SD=0.776)           & t=2.584  & 0.008   \\
                                       & Negative & M=2.833 (SD=2.531)           & 
                                       M=3.833 (SD=3.522)*       
                                       & t=-1.846 & 0.039  \\

\bottomrule
\end{tabular}
\label{Table:InterT}
\end{table*}

\subsubsection{Behavioral Engagement Results}
In general, children were more behaviorally engaged with the MR games than with the traditional video games. Children exhibited more positive behaviours in the MR games (p=.009,t=2.526) and more negative behaviours in the traditional video games (p=.003,t=-3.028). When playing as a "supplier", more negative behaviours were seen during traditional video games (p<.001,t=-3.506). Similarly, when children played as a "builder", the MR games showed more positive behaviours (p=.005,t=2.782), while the traditional video games showed more negative behaviours (p=.033,t=-1.937). But there was no statistically significant difference that appeared for positive behaviours in different games while playing as a "supplier"(p=.308,t=.509).

\subsubsection{Emotional Engagement Results}
Children that participated in the MR game showed higher levels of emotional engagement. The results suggested that children had more significant positive emotions (p<.001,t=4.240) when playing MR games, and significantly more negative emotions (p<.001,t=-4.031) in the traditional video games.  While playing as a "builder", more positive emotions (p=.020,t=2.172) were found during the MR game, and more negative emotions (p=.052,t=-1.696) were seen during the traditional game. Moreover, when playing as a "supplier", the traditional video game results in more negative emotion(p=.005,t=-2.782). There was no statistically significant positive emotion engagement difference between the two games while playing as a "supplier"(p=.186,t=.912).

\subsubsection{Verbal Communication Results}
Compared with the traditional video game, children had more positive verbal communication in the MR game, as evidenced by a greater number of terms relating to positive collaboration (p=.008, t=2.584). When children performed the role of the builder, there were more positive words in the MR game (p<.001,t=3.498) and more negative words in the traditional video game (p=.005,t=-2.814). There was no statistically significant difference between the two games when children played as the "supplier".
 
\subsection{Emotional Perception Experience}
We analyzed children's emotional perception experiences in both games using interview data and found that both games enhanced children's sense of co-presence and strengthened their connection with their companions. In both games, the majority of children reported feeling in the same room or face-to-face with their partners: 87.5\% in the MR game and 79.2\% in the traditional video game. However, there was no substantial difference between the two games in terms of how children perceived co-presence.  In addition, both games promoted children's closeness to others, as shown by 91.7\% in the MR game and 87.5\% in the traditional video game. Likewise, the children's sense of closeness to their partners did not show a significant difference between the two games.

\subsection{Preference and Reasons}
Children reported their preference for both games after completing the experiment, and we found that the MR game were superior to the traditional video game in both overall preference and preference for interaction. In terms of overall game preference, 54.2\% of children preferred the MR game and 37.5\% preferred the traditional video game. As for the interaction, 54.2\% of children preferred to interact with the toolkit in the MR game and 33.3\% preferred to interact with the mouse in the traditional video game. Children who preferred the MR game generally reported that the interaction of the MR game was more interesting and closer to reality, whereas using mouse for control were more difficult to handle precisely.. For example, P2 said:
\begin{quote}
   \emph{"MR game is really like building LEGO bricks, and traditional video game is more like playing games, although I know they are both games."}
\end{quote}
Children who preferred the traditional video game reported that was simpler to use a mouse, and the MR game was too complicated to interact with through the toolkit. For example, P20 said:
\begin{quote}
   \emph{"It takes too much time for me to find cards, but it's very convenient to click with a mouse."}
\end{quote}
Additionally, we asked the children which game they felt was harder to play, with 50\% responded the MR game and 33.3\% answering traditional video game, the rest suggested both games have same difficulty. We then used correlation analysis to verify the corelation among children's familiarity with mouse control and overall preference of games, preference for interactions, and attitudes toward difficulty of the interaction. The results showed there is no statsically signifiant correlation between the mouse control familiarity and the other three evaluations. However, under the same mouse control familiarity, the overall preference and the preference of using the MR toolkit shows a positive correlation($\gamma$= .486), and a weak negative correlation between the overall preference and attitude toward difficulty ($\gamma$=-0.220).

\section{Discussion}
\subsection{Effects on improving social and collaborative skills and alleviating social loneliness}
Overall, based on the results of quantitative and qualitative analysis, our system achieved the intention of improving children's social and collaborative skills. In the game, most of the children realized the importance of collaboration and developed a sense of teamwork. Recognizing the value of collaboration can help overcome self-centred problems in young children's development~\cite{dongsheng2021probe}. 

\begin{quote}
   \emph{"It is a team effort. We should cooperate to support and correct each other."}
\end{quote}
P1 reported this when being asked about "What's the most important thing you learned from this game?", and 10 other children mentioned "collaboration" or "teamwork" to this question.
\begin{quote}
    \emph{"I learned to be patient if my partner is slow...I was a little upset at first and then I slowly got it under control, because I want everyone to have fun. I'm usually very easy to get angry, but this time I controlled myself."}
\end{quote}
P12 reported this to the same question, and 3 other children mentioned "be patient" as well. Patience is intimately related to self-control and is essential for children to participate in cooperation \cite{alan2018fostering}. In addition to the teamwork and patience mentioned above, many children highlighted being considerate of others, completing their own tasks promptly, avoiding causing trouble for their partners, understanding and assisting each other, and discussing calmly when encountering difficulties, etc. Indeed, social and collaborative skills are originally composite structures covering multiple skills, such as communication, team support, and problem-solving \cite{graesser2018advancing}. The specific insights that children learn during the collaborative process serve as the foundation for their eventual growth of generalized social and collaborative skills. According to the interview, we found that they gained a more precise and deeper understanding of collaboration and mastered certain approaches to collaboration. 

We also discovered that using Mr. Brick system can help alleviate children's social loneliness. Slavin et al. \cite{slavin2013cooperative} suggested that cooperation not only improves children's cognition and performance, it also shows great potential to promote social-emotional outcomes. We met a touching case in the experiment, P9, a very introverted six-year-old girl. When she came to our experiment, she held her mother's hand tightly and hid behind her mother's back, holding an reversible octopus plushie in her hand. This plushie was originally designed for children with autism who have difficulty expressing their emotions. The plushie's pink side indicates happy feelings, while the blue side represents negative emotions. P9 was the only one of our participants who asked for the parent to be in the experimental room. Before the experiment, P9 barely spoke and did not dare to make eye contact with the experimenter. Her octopus plushie was switched to the blue side. 

\begin{figure*}[h]
    \centering
    \includegraphics[width=\textwidth]{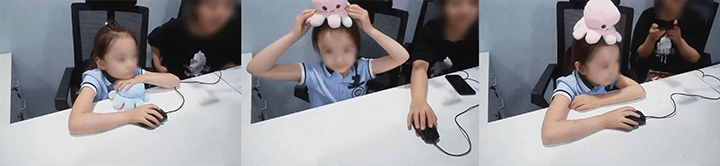}
    \caption{P9 and her reversible octopus plushie}
    \label{fig:P9}
\end{figure*}

As the game progressed, we were delighted to observe some changes. First, we noticed that her facial expression had softened, gradually changing from expressionless to frequently smiling. The frequency of her verbal requests grew, as did her tone and volume. Her verbal interactions with her partner have progressed from simple responses to rich conversations that include descriptions, requests, reminders, instructions, enquiries, etc. Surprisingly, during the break, P9 said that she can complete the game independently without the company of her mother. In the next round, P9 changed the octopus plushie from blue to pink and placed it on her head (Fig.~\ref{fig:P9}). This was a significant positive sign that she was gradually becoming more socially active during the game. 

Apart from P9, most children reported feeling more connected to their partners during the game. In addition, when asked whether this game can help them relieve loneliness and maintain a pleasant mood during the pandemic, the majority of the children responded positively. Compared with adults, children and adolescents are at higher risk of mental health problems when they are forced to reduce social activities during the epidemic \cite{zhou2020managing}. According to early indicators of COVID-19, more than a third of children and adolescents report high levels of loneliness \cite{loades2020rapid}. Numerous studies \cite{jiao2020behavioral, meherali2021mental} indicated that maintaining social connections with peers can help children through this difficult time, including organizing support groups, keeping regular social interactions, such as video calls; and playing collaborative games. Our games have demonstrated the potential to be a tool for assisting children in alleviating social loneliness.

\subsection{Differences between MR game and traditional video game}
Many of the differences between the MR game and the traditional video game were also found in interviews. 

According to children, the biggest advantage of the MR game is \emph{"a sense of reality"}(P21). Many children reported that the MR game gave them a more physical experience than the traditional video game, \emph{"like actually building LEGO bricks,"}, P2 said. MR game allows direct interaction with physical objects and direct observation of the player's hand movements, which is more \emph{"simple to learn"}(P1, P19) and \emph{"more realistic"} (P19, P21). These encouraging remarks from the children on the MR game further validated our design guideline 3, and offered evidence for past studies' conclusions that augmented manipulatives with conceptual metaphors based on image schemas should facilitate learning\cite{marichal2017ceta,antle2013getting}. 

\begin{figure*}[h]
    \centering
    \includegraphics[width=\textwidth]{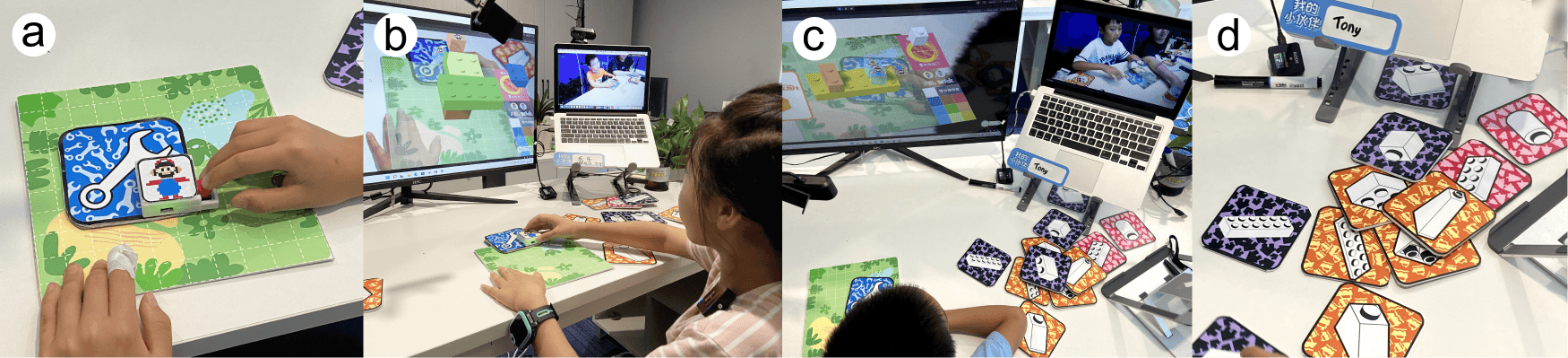}
    \caption{Details of children's manipulating TUI: a) a closer look at the tangible clicker and the board during children's manipulation; b) a girl moving the clicker on the board to build a virtual brick; c) a boy's perspective of the mixed-reality game as a "builder"; d) the colorful cards with recognition patterns.}
    \label{fig:manipulate}
\end{figure*}

Fontijn et al. demonstrated that an interface that is physically engaging enhances fun\cite{fontijn2007functional}. In our study, MR game was appraised by children as \emph{"more interesting"} (P7) and \emph{"more fun"} (P16). Several participants also note the card's unique tactile feel. For example, P4 mentioned\emph{"The sensation of the board's attachment to the table reminds me of ice skating, which I find quite enjoyable."}. P16 reported that \emph{"It was fun having all the cards lined up in your hand, and I like the beautiful colors of the cards."} (see Fig.~\ref{fig:manipulate}). We can infer that the hybrid setup of MR.Brick opens up their tactile perspective of sensory, which enhances their game experience. This is analogous to zhou et al.'s Magic Story Cube\cite{zhou2004magic}, which simultaneously makes the multi-sensory experience more interesting by enhancing the feeling of physical touch.

According to Fontijn et al. \cite{fontijn2007functional}, getting a sense of accomplishment is the first core source of fun, which is determined by the balance between challenge and control. While the challenge offered is generally seen as the most important factor in making a game enjoyable\cite{malone2021making}.
In our study, several children mentioned that the MR game gave them a \emph{“sense of control”} by using the board to rotate and observe, and using cards to manipulate and select. Here are two examples:
\begin{quote}
    \emph{"It(MR game) gives me the feeling that I am the captain of the train, which is very pleasant, and everything is under my command."}(P4)
    
    \emph{“The traditional video game is all about the electronic element on the screen. But if it's the MR version, you have a projector that you can look at, you know, you can take full control of the game, and it's like I'm in control of the world.”(P11)}
\end{quote}
At the same time, the MR game was reported to be \emph{"a little challenging"}(P6, P22) and \emph{"Have a sense of accomplishment when completing hard mode"}(P8). Hence, we made a great balance between challenge and control by offering a sense of control to the player, but at the same time difficult enough for the outcome to be uncertain~\cite{fontijn2007functional}. Our findings support the conclusions of many previous studies~\cite{manches2011designing,davis2015spatial,yannier2016adding} that demonstrated experimenting with three-dimensional physical objects in mixed-reality situations results in more learning and satisfaction than two-dimensional interaction on a flat screen. 

However, compared to the traditional video game, children reported that the MR game caused them to spend more time looking for the correct card than playing the game(P12). Similarly, P20 reported that \emph{"finding the right card"} is the most difficult thing during the game. Possibly owing to the large quantity and size of the cards, children were unable to lay out and see the patterns on all the cards throughout the experiment.

Some children also cited traditional video game benefits that MR games lack.
\emph{"You don't need a lot of props."}, P8 commented, thinking it is not convenient if players need to prepare a lot of extra cards to start the game. Also, some children who were more familiar with mouse control said that the traditional video game was more convenient than the MR game. For example:
\begin{quote}
    \emph{"I used to take a programming class, so I became fairly proficient with the mouse by holding it like this and dragging it. This (MR game) must be picked up and moved to its proper location. Then this (traditional video game) is pretty easy; simply click the mouse, and that's all that is required."}(P24)
\end{quote}
But not all children found the traditional video game easier to play, and some younger children reported that the mouse was harder to use(P1, P2, P8, P9, P10, P16, P19, P23). We did not find significant differences in the rating of operation difficulty and preferences from the interview feedback. According to Yannier et al.~\cite{yannier2016adding}, for students’ learning, there were no significant differences between the mouse control and the tangible trigger, and the effect of observing physical phenomena was more powerful than the effect of using a simple tangible trigger. Therefore, we assume that ratings of the difficulty may mainly be related to children's age and familiarity with mouse control.

Using a mouse to determine the correct position in a 3D environment proved difficult for the vast majority of children. For example, \emph{“I've already clicked on that spot, but he doesn't move, which is quite unpleasant.”}, P11 argued. This is mostly owing to the limited spatial awareness skills of our participants~\cite{rosser1994cognitive}. They are unaware that while viewing the model through the camera in 3D software, the model will have a perspective effect and the actual location of the assembly will be somewhat behind the position of the mouse click. This may explain why their average time spent playing traditional video games was less for simple games than for MR games and longer for sophisticated ones.

\begin{figure*}[h]
    \centering
    \includegraphics[width=\textwidth]{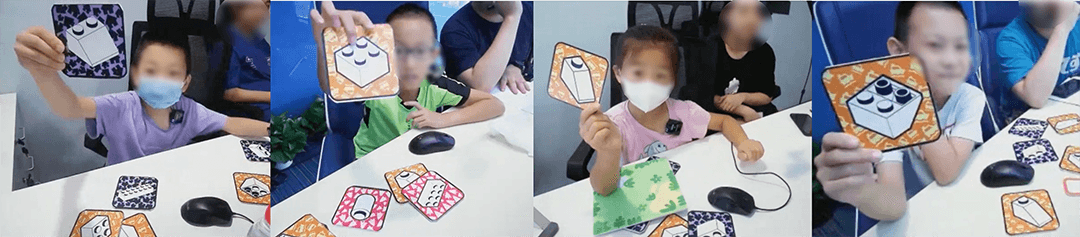}
    \caption{Children using tangible cards to describe their needed piece during the traditional video game.}
    \label{fig:my_label}
\end{figure*}
Another interesting finding is that during the traditional video game, in order to improve communication efficiency and achieve better collaboration, many children prefer to use tangible props to make up for the limitation in the verbal description, even when we strongly suggest they try not to use the props.\emph{“It might take 20 seconds to describe. But only two seconds with this(card) .”}, P24 told us. 
\begin{quote}
    \emph{“It's hard to tell what this part looks like. It's better to use cards. With props, I don't have to talk much. It feels more convenient. All you need to do is to pick up the card.”} (P21)
\end{quote}
Kuzuoka et al.~\cite{kuzuoka1992spatial} indicated that gestures significantly increase communication efficiency and substantially reduce the number of verbal expressions required to indicate intention when using video conferencing to convey gesture information for remote collaboration. Because children's verbal abilities are limited, their subconscious choices provide solid evidence that the mixed-reality with TUI approach is more child-friendly, and it increased the efficiency of our young participants' remote collaboration by allowing them to employ gestures and props.
\subsection{Design implications}
Looking back through our design guidelines based on our experimental results, we found a number of design opportunities worth promoting in the future, as well as some technical and design considerations.
\subsubsection{\textbf{The mixed-reality setup}} It is enticing to children and can improve their manual dexterity and remote collaboration~\cite{campos2011fostering}, making the game more engaging~\cite{aslan2019investigating} and boosting the overall gaming experience. Similarly to what yuan et al. suggested~\cite{yuan2021tabletop}, we propose that designers can try linking the physical object of each distant player to generate a greater sense of co-presence, or they might investigate whether there are more intriguing uses for physical objects in remote collaboration than as game manipulation tools.

\subsubsection{\textbf{Game mechanics that encourage collaboration}} Through the right arrangement of mutual confinement and interdependence of game rules, a collaborative environment may be fostered\cite{antle2013getting}. According to the children's feedback, the design of both roles in our game is very interesting, and regardless of which role they play, nearly all of the children report a self-contribution of 50\%, indicating that our game rule design encourages the players to achieve an interdependent but equal collaborative relationship. According to Woodward et al.~\cite{woodward2018investigating}, providing equal opportunities for interaction is important for collaborative tasks, otherwise children may have negative behaviours towards other members.  Thus, it is important for designers to consider equal contributions from each player and create an equal collaborative environment when designing rules. 
\subsubsection{\textbf{System usability}}
Since MR content recognition is unstable and occasionally lost, it is necessary to maximise the size of the recognition pattern. For instance, in order for our cards to be more easily identified by the camera, we raised their size. Nonetheless, it was difficult to locate cards throughout the game owing to the enormous amount of cards. The designers of MR must create a balance between recognizable sizes and child-friendly operational sizes.
\subsubsection{\textbf{Children-friendly mixed-reality 3D game}}
By employing independent cameras and screens, we were able to circumvent the physical location difficulties and perspective distortion problems that children encountered in prior research~\cite{kang2020armath}. The addition of a virtual cursor compensated for children's deficiencies in hand-eye coordination and verbal description~\cite{dunser2007lessons,kang2020armath}. The rules and complexity of our game are currently suitable for children, however, the 3D operation part may be improved. The spatial perceptual limitations of children should be taken into mind~\cite{rosser1994cognitive}. 3D issues are almost unavoidable in mixed reality systems, therefore designing 3D games for children based on MR technology is a topic that merits more research and consideration. Children-friendly 3D model operation in video games will be one of our future endeavours.
\subsubsection{\textbf{Adapted scaffold}}
Mr. Brick, our agent, may lead and encourage children to communicate more and guide the game's advancement, but in the real situations, children still require the assistance of researchers. The agent is inefficient as an adapted scaffold for guiding children's play. An adaptable scaffold can assist children in achieving a good learning state~\cite{azevedo2004does,hourcade2008interaction}, establishing a balance and developing an adaptable scaffold needs effort.

\subsection{Limitation and future work}

For the MR. Brick system, it still has some technical limitations as a proof-of-concept prototype, mainly the equipment required to access the game settings. By removing the restriction in the augmented reality algorithm that necessitates a perpendicular angle for accurate position calculation, it will be possible to reduce the number of cameras to one. The number of screens required can also be reduced to one by integrating the video communication tool into our game. Furthermore, tangible cards can be produced simply by printing, making the game more accessible. 
In addition, it is important to encourage children's creativity by providing opportunities to build freely and adapted to their ability during the game. According to Resnick and Silverman~\cite{resnick2005some}, “Low Floors, High Ceilings, and Wide Walls” should be followed when developing tools aimed at facilitating children's learning and development through making. That is to say, successful tools allow new users to adapt easily, allow more experienced users to develop complex structures, and allow users to create freely. Mr. Brick has the potential to continue to be developed into a more complete educational tool.
With the elimination of these limitations, Mr. Brick may be more simply republished on portable devices using the Unity framework, opening up more opportunities for further research or actual marketing.

For the study, although our research has confirmed that Mr. Brick has the effect of developing children's social and collaborative skills, however, children only played the MR game for 20-30 minutes in our study, which was a short-term experience. It is difficult to say whether this is due to novelty effects. Besides, the game used in our study only contains a small number of relatively simple and fixed game tasks. In the future, we can explore whether this kind of game can improve children's social and collaborative skills based on long-term research. For example, a semester-long weekly experiment in an elementary school class.
Moreover, more measurement such as network satisfaction survey can be involved in the study to help further research in the social communication behaviour of children.

\section{Conclusions}
This paper presents MR.Brick, a mixed-reality remote collaboration system for children aged 6 to 10 that intends to foster children's social and teamwork abilities by playing collaborative brick-building activities. A controlled experiment with 24 children was conducted. Based on the results of quantitative and qualitative analysis, it has been demonstrated that MR.Brick achieves its intended purpose of enhancing children's social and collaborative skills. Besides, MR.Brick with MR and TUI increased children's emotional and behavioural involvement in the remote collaborative game compared to the traditional video version. We anticipate that this work will contribute to the educational technology research community by offering a unique interactive system and a thorough knowledge of distant cooperation for children.


\begin{acks}
This project is supported by National Natural Science Foundation Youth Fund 62202267. Our sincere thanks to all the children of our experiment. 
\end{acks}

\bibliographystyle{ACM-Reference-Format}
\bibliography{sample-base}

\appendix

\section{Experiment Details}

\subsection{Demographic information of participant children}

\begin{table*}
\caption{Demographic information of the child participants}
\label{tab:demographics}
\begin{tabular}{llllll}
\hline
Group ID & Child ID & Age & Gender & \begin{tabular}[c]{@{}l@{}}Familiarity with mouse control\end{tabular} & \begin{tabular}[c]{@{}l@{}}First\\ Game Type\end{tabular} \\ \hline
1        & P1       & 8   & F      & 3                                                                         & MR                                                        \\
1        & P2       & 9   & M      & 4                                                                         & MR                                                        \\
2        & P3       & 10  & M      & 4                                                                         & Traditional                                               \\
2        & P4       & 8   & M      & 5                                                                         & Traditional                                               \\
3        & P5       & 9   & M      & 4                                                                         & Traditional                                               \\
3        & P6       & 8   & F      & 5                                                                         & Traditional                                               \\
4        & P7       & 7   & M      & 3                                                                         & MR                                                        \\
4        & P8       & 8   & F      & 4                                                                         & MR                                                        \\
5        & P9       & 6   & F      & 3                                                                         & MR                                                        \\
5        & P10      & 10  & F      & 2                                                                         & MR                                                        \\
6        & P11      & 8   & M      & 3                                                                         & Traditional                                               \\
6        & P12      & 8   & M      & 5                                                                         & Traditional                                               \\
7        & P13      & 7   & F      & 5                                                                         & Traditional                                               \\
7        & P14      & 9   & M      & 5                                                                         & Traditional                                               \\
8        & P15      & 7   & M      & 3                                                                         & MR                                                        \\
8        & P16      & 6   & F      & 2                                                                         & MR                                                        \\
9        & P17      & 9   & M      & 4                                                                         & MR                                                        \\
9        & P18      & 9   & F      & 3                                                                         & MR                                                        \\
10       & P19      & 8   & M      & 5                                                                         & Traditional                                               \\
10       & P20      & 10  & M      & 5                                                                         & Traditional                                               \\
11       & P21      & 8   & M      & 5                                                                         & Traditional                                               \\
11       & P22      & 8   & M      & 3                                                                         & Traditional                                               \\
12       & P23      & 8   & M      & 5                                                                         & MR                                                        \\
12       & P24      & 8   & M      & 4                                                                         & MR                                                        \\ \hline
\end{tabular}
\end{table*}

\subsection{Online questionaire}
\textbf{Investigation on Children's Social Collaboration Ability and Emotional State during Epidemic
}

Hello, Parents! We are developing remote systems to mitigate the impact of social isolation on the mental health of children during the epidemic.
The main purpose of the questionnaire was to find out the children's preference for constructive games, their social cooperation ability and their emotional status in the epidemic environment. Thank you very much for participating in this questionnaire. All the answers are only for statistical analysis and academic research. Please fill in the questionnaire carefully according to your own actual situation and answer with confidence! If you and your child are interested in participating in the experience of our building system and becoming our Little Experience Officer, please leave your contact information at the end of the questionnaire.

\subsubsection{What is your child's age?
(If your child's age is not in the range given by the questionnaire, you do not need to fill in this questionnaire. Thank you for your cooperation!)}
\begin{itemize}
    \item 12 years old
    \item 11 years old
    \item 10 years old
    \item 9 years old
    \item 8 years old
    \item 7 years old
    \item 6 years old
\end{itemize}

\subsubsection{What is your child's gender? }
\begin{itemize}
    \item male
    \item female
\end{itemize}

\subsubsection{Has your child played the following construction games? [multiple choice]}

\begin{itemize}
    \item No
    \item Solid toys: such as Lego blocks
    \item Virtual games such as My World
    \item Other Construction Games
\end{itemize}

\subsubsection{Please assess your child's familiarity and preference for building games such as Lego Blocks and My World: [Matrix Questionnaire] (1:Bad 2:Almost 3:Good)}
\begin{itemize}
    \item Very interested in this type of game
    \item Familiarity with building structures such as building blocks
    \item Read Lego spelling instructions and assemble independently
    \item Skilled in collecting and using different parts
\end{itemize}

\subsubsection{Please assess if your child's mood or state increases during the epidemic: [Matrix Questionnaire] (Note that all of the following mood or state assessments are compared with the pre-epidemic child's state. If there is an increase in mood or state, rate the increase by 2-5 depending on the magnitude of the increase. If it remains exactly the same or even weakens before the epidemic (regardless of mood or state), fill in 1.
(1:It fits perfectly; 5:Very inconsistent)}
\begin{itemize}
    \item Because of prolonged isolation at home, children may feel a little uncomfortable and sometimes anxious
    \item Children have nightmares because they have been quarantined at home
    \item Children are more often sleepy than they were before the epidemic and feel less energetic at home
    \item It's easier to pick food when you're away from home than before
    \item It's easier to wake up at night than before, or a little sleepless
    \item Some fears and concerns about the health of relatives, such as nucleic acid status
    \item Forced to request news updates (such as news, daily events, etc.) and keep asking for information about the epidemic
    \item Compared to before the epidemic, it feels like your child is sometimes a little worried, as if he was worried
    \item Responses to everyday incidents and mood swings seem to be worse than before quarantine
    \item During home isolation, it's easier to skip classes, homework, etc.
    \item Feels more clingy than it was before the epidemic
\end{itemize}

\subsubsection{How many good friends does your child have? [Drop-down questions]}
\begin{itemize}
    \item No
    \item 2 to 3
    \item 4 or more
\end{itemize}

\subsubsection{How many times a week does your child play with these friends? [Drop-down questions]}
\begin{itemize}
    \item Less than once
    \item 1 to 2 times
    \item 3 or more times
\end{itemize}

\subsubsection{How do your children behave compared to their peers in the following areas? [Matrix Questionnaire] (1:Bad 2:Almost 3:Good)}
\begin{itemize}
    \item Getting along with other children
    \item Attitude to Parents
    \item Learn and play by himself/herself
\end{itemize}

\subsubsection{Next, guide your child to rate the following three questions. Score 1-3 from disagreement to disagreement: [Matrix Questionnaire] (Note that the following three questions do not need to be compared with the pre-epidemic situation, but only record the child's current state. Please guide the child to fill out his or her own feelings in a relaxed and safe environment.)(1:It fits perfectly; 3:Very inconsistent)}
\begin{itemize}
    \item Do you often feel like you're out of company?
    \item Do you often feel left out in the cold?
    \item Do you often feel isolated?
\end{itemize}

\subsubsection{Next, guide your child to rate the following three questions. Score 1-3 from very inconsistent to very consistent:(1:It fits perfectly; 3:Very inconsistent) [Matrix Questionnaire]}
\begin{itemize}
    \item When I work with other children to complete tasks or games, I often take the initiative to share ideas, information and resources.
    \item When I work with other children to complete tasks or games, I always participate with an open mind in solving problems collectively and sharing ideas and ideas without hindering others' contributions.
    \item I always listen to, respect, recognize and support others’efforts when I work with other children to complete tasks or games.
\end{itemize}

\subsubsection{Thank you very much for your participation! Do you agree that we will invite you and your children to XXX Institute to participate in the building game experience? (Small gifts will be given to you after our event ends) [Radio Title] }
\begin{itemize}
    \item yes
    \item no
\end{itemize}

\subsubsection{Are you in XXX city? Can you participate offline? [Radio Topic]}
\begin{itemize}
    \item yes
    \item no
\end{itemize}

\subsubsection{Please leave your mobile number: [Single line text title]}

\subsection{Adapted SUS scale}

(five-Likert scale, 1:It fits perfectly; 5:Very inconsistent)
\begin{itemize}
    \item I think I'll play this game a lot
    \item I think this game is too difficult
    \item I think this game is too easy
    \item I need help to play this game
    \item I think the game is well designed and both roles are fun to play
    \item I think my classmates can also learn this game quickly
    \item I am confident that I can do well in this game
    \item I need to know a lot of other knowledge and information in advance to play this game
\end{itemize}

\end{document}